\documentclass[10pt,final,journal,twocolumn]{IEEEtran}
\usepackage{cite}
\usepackage[cmex10]{amsmath}
\usepackage{graphicx,epsfig,amssymb}
\usepackage{anysize}
\usepackage{graphicx,cite,epsfig,subfigure}
\usepackage{amsmath,amssymb,geometry}
\usepackage{amsmath,bm}
\usepackage{setspace}
\usepackage{multirow}
\usepackage{booktabs}
\usepackage{amsmath}
\usepackage{xcolor}
\usepackage{float}
\marginsize{0.7in}{0.7in}{0.8in}{1.2in}
\begin{document}

\title{
\textsc{5G ultra-dense networks with non-uniform distributed users}}
\vspace{0.1cm}
\author{\normalsize
Junliang Ye$^1$, ~\IEEEmembership{Student~Member,~IEEE}, Xiaohu Ge$^1$, ~\IEEEmembership{Senior~Member,~IEEE}\\
Guoqiang Mao$^2$,~\IEEEmembership{Senior~Member,~IEEE}, Yi Zhong$^1$,~\IEEEmembership{Member,~IEEE}\\

\thanks{\small{Submitted to IEEE Transactions on Vehicular Technology.}}
\thanks{\small{Correspondence author: Dr. Xiaohu Ge,  Tel: +86 (0)27 87557941, Fax: +86 (0)27 87557941
Email: xhge@mail.hust.edu.cn.}}
\thanks{\small{The authors would like to acknowledge the support from the NFSC Major International Joint Research Project under the grant 61210002, National Natural Science Foundation of China (NSFC) under the grants 61301128 and 61461136004, the Ministry of Science and Technology (MOST) of China under the grants 2014DFA11640 and 2012DFG12250, the Fundamental Research Funds for the Central Universities under the grant 2015XJGH011, the Special Research Fund for the Doctoral Program of Higher Education (SRFDP) under grant 20130142120044. This research is partially supported by the EU FP7-PEOPLE-IRSES, project acronym S2EuNet (grant no. 247083), project acronym WiNDOW (grant no. 318992) and project acronym CROWN (grant no. 610524).}}
}
\maketitle

\markboth{IEEE Transactions on Vehicular Technology, Vol. XX,
No. Y, Month 2017} {Ge etc.: Performance analysis of 5G ultra-dense small cell networks based on Radiation and Absorbing models\ldots}

\date{\today}
\renewcommand{\baselinestretch}{1.2}
\thispagestyle{empty} \maketitle \thispagestyle{empty}
\newpage
\setcounter{page}{1}\begin{abstract}

User distribution in ultra-dense networks (UDNs) plays a crucial role in affecting the performance of UDNs due to the essential coupling between the traffic and the service provided by the networks. Existing studies are mostly based on the assumption that users are uniformly distributed in space. The non-uniform user distribution has not been widely considered despite that it is much closer to the real scenario. In this paper, Radiation and Absorbing model (R\&A model) is first adopted to analyze the impact of the non-uniformly distributed users on the performance of 5G UDNs. Based on the R\&A model and queueing network theory, the stationary user density in each hot area is investigated. Furthermore, the coverage probability, network throughput and energy efficiency are derived based on the proposed theoretical model. Compared with the uniformly distributed assumption, it is shown that non-uniform user distribution has a significant impact on the performance of UDNs.

\end{abstract}
\newpage
\IEEEpeerreviewmaketitle \vspace{-1cm}

\section{Introduction}
\label{sec1}
\subsection{Related works}

The fifth generation (5G) mobile communication systems are envisaged to provide a 1000 times enhancement of the network capacity while achieving a much higher energy efficiency compared with the fourth generation (4G) mobile communication systems. The ambitious aims of 5G mobile communication systems bring both opportunities and challenges to researchers all over the world \cite{Demestichas13}. The UDNs are regarded as one of the key technologies for 5G mobile communication systems \cite{Andrews14}. The main difference between UDNs and heterogeneous networks (HetNets) lies in the dramatic increase of small cell base station (SBS) density. The distances between users and SBSs are greatly reduced with the increase of the SBS density, hence more wireless links are available for users in wireless networks to enhance the quality of service (QoS) \cite{Ge16}. On the other hand, UDNs also suffer from the increasing energy consumption with the massive deployment of SBSs. Therefore, one of the core problems for deploying UDNs is the optimization of SBS density to meet the traffic demand in an energy efficient way in hot spot areas.

 Some recently studies were conducted to evaluate the performance of UDNs. A basic question was investigated in \cite{Tu16}: how to choose the SBS density for a 5G UDN when the backhaul capacity and energy efficiency are jointly considered. A comprehensive introduction of researches about the energy efficiency of UDNs was given by G. Wu et al. and the tradeoff between energy efficiency and spectrum efficiency was highlighted in \cite{Wu15}. A resource allocation scheme was proposed to analyze trade-offs between the spectrum efficiency and energy efficiency of 5G UDNs  \cite{Wu15}. The influence of SBS density on the outage probability was discussed and different multiple access technologies were compared to obtain the optimal solutions for allocating subcarriers with interference constraints \cite{Stefanatos14}. The Lyapunov method and mean field game were utilized to optimize the energy efficiency with the QoS constraints \cite{Samarakoon16}. The impact of the spectrum bandwidth and SBS density on the capacity and the spectrum efficiency was analyzed in \cite{L車pez-P谷rez15}. Comparisons between different spectrum bandwidths and SBS densities were made by D. L車pez-P谷rez et al. to propose a technique for increasing the average user capacity to 1 gigabit per second (Gbps) \cite{L車pez-P谷rez15}. Cognitive radio technologies were adopted to enhance the throughput of 5G UDNs \cite{Tseng15}, and energy efficient optimal power allocation strategies subject to constraints on the average interference power are proposed in \cite{Zhou16}. On the other hand, the stochastic geometry theory is widely used to evaluate the performance of UDNs. The delay of heterogeneous cellular networks with spatial temporal traffic was investigated by using stochastic geometry in \cite{Zhong17}. Closed-form formulas of network capacity and energy efficiency were derived in \cite{Zhang16} by modeling the wireless network using the stochastic geometry tools. The difference between line of sight (LoS) channel and non-line of sight (NLoS) channel was considered to evaluate the impact of channel fading on the spatial spectrum efficiency \cite{Ding16}. It was shown in \cite{Ding16} that the network spatial spectrum efficiency will not monotonously increase with the increase of the SBS density when LoS and NLoS channels are considered. A new type of UDNs including femtocell base stations (FBSs), macro cell base stations (MBSs) with distributed antennas was proposed in \cite{Yunas15}, where the energy efficiency, spectrum efficiency and spatial spectrum efficiency of UDNs were improved. By utilizing an innovative millimeter-wave (mmWave) decoupling method, mmWave can be used by network users for uplink transmissions to traditional microwave base stations (BSs) \cite{Park16}. Based on the mmWave decoupling method, a closed-form formula for spectrum efficiency of UDNs was derived and a resource management method was proposed to optimize downlink transmission rate of UDNs with a minimum uplink rate constraint. The backhaul traffic of UDNs based on the mmWave and massive MIMO technologies was analyzed and the impact of different pre-coding methods was also investigated in \cite{Gao15}. When both the long-term evolution (LTE) and the wireless fidelity (WiFi) were deployed in UDNs, a Markov chain model was utilized to analyze the performance of the proposed LTE-WIFI UDNs \cite{Galinina15}. Based on the energy harvesting technology, the trade-off between the energy efficiency and the QoS was analyzed for UDNs \cite{Ghazanfari16}. Furthermore, a fractional programming method was adopted to investigate the energy efficiency maximization problem of wireless networks under a minimum system throughput constraint \cite{Wu16}.

 Though the aforementioned studies have investigated many aspects of UDNs, most of them only focused on studying the influence of SBS density on the capacity and energy efficiency. Few effort has been spent on the effect of the SBS density on both the network performance and the user experiences simultaneously. Recently, the virtual cell technology has been widely used to improve the user experience of UDNs \cite{Chen16,Kim14,Wang16,Nie16,Hong13,Li15}. By adopting the virtual cell technology, a new type of UDNs termed user-centric UDNs were introduced in \cite{Chen16}. The main idea of user-centric UDNs is to cluster SBSs dynamically and intelligently to provide better services for users. Based on virtual cells in UDNs, a new type of beam forming technology named the balanced beam forming algorithm was developed to optimize the network capacity \cite{Kim14}. A type of virtual cell networks based on distributed antennas were presented in \cite{Wang16}, and the influence of cell size on the users' maximum downlink achievable rate was investigated. Based on stochastic geometry theory, the trade-off between the energy efficiency and spectrum efficiency of UDNs was investigated to optimize the energy efficiency with the minimum network capacity constraints \cite{Nie16}. By the joint optimization of virtual cell clustering method and the beam forming algorithm, the sum capacity was improved for UDNs \cite{Hong13}. By optimizing the interference nulling range, the user-centric interference nulling method was proposed in \cite{Li15}, where the outage probability is reduced by 35\%--40\%.

\subsection{Main contributions}

From the above discussion, one can observe that the focus of the researches on UDNs turns from network performance to user experience. Moreover, the design and optimization of UDNs develops from network-centric to user-centric. Due to uncertainties of users' activities, few work have studied the impact of users' activities on the network performance . By measuring the entropy of each individual's moving trajectory, Song et al. found a 93\% potential predictability in user mobility \cite{Song101}. Based on the human mobility trajectory measured from real data, the individual mobility model considering the human mobility tendency habit was proposed to describe the human mobility in the real world \cite{Song102}. The gravity law model was established to analyze the number of commuters between different areas \cite{Zipf46}. The theoretical result of \cite{Zipf46} was improved in \cite{Simini12}. By proposing the Radiation and Absorbing (R\&A) model and comparing it with the empirical data, the R\&A model was proved to be highly reliable in large range of spatial scale.
Although there exist a large amount of meaningful and important studies for UDNs, one basic question is still not well answered: how many SBSs are required to meet the greatly increased traffic demand while guaranteeing a high energy efficiency? To answer this question, a theoretical model including user density, BS deployment, interference analysis and energy consumption need to be established for 5G UDNs. Based on this unsolved problem, the contributions and novelties of this paper are summarized as follows.

\begin{enumerate}
\item Based on the queueing network theory, the R\&A model is first used to analyze the effect of the user density on the performance of 5G ultra-dense small cell networks in the temporal domain.
\item By utilizing the integral geometry and Euler summation, the impact of user and SBS densities on the coverage probability and average number of SBSs in virtual cells is investigated. Furthermore, by adopting a two-dimensional Markov chain, the blocking probability and channel occupancy rate are analyzed for 5G ultra-dense small cell networks based on the R\&A model.
\item Based on the theoretical analysis on channel capacity and energy consumption of 5G ultra-dense small cell networks, the network energy efficiency is investigated using the proposed models. Simulation results indicate that the non-uniformly distribution of users has a profound effect on the energy efficiency.

\end{enumerate}

The rest of this paper is organized as follows. Section II describes the system model of 5G UDNs. The R\&A model is introduced in section II-A, then the model is extended by the queueing network theory in section II-B. The interference model and the SBS service model are described in section II-C and section II-D, respectively. Network performance metrics are analyzed and derived in Section III. Numerical and simulation results are displayed in section IV. Finally, Section VI concludes this paper.

\section{System Model}
\label{sec2}
\subsection{Radiation and Absorbing model}
Definitions of some default parameters are shown in the following table.

\begin{table}[!htbp]
\renewcommand{\arraystretch}{1.3}
\caption{An Example of a Table}
\label{table_example}
\centering
\begin{tabular}{|c|c|}
\hline
${\mathrm{HS_i}}$ & Hot spot $i$ \\
\hline
${\mathrm{CR_i}}$ & Coverage region of ${\mathrm{HS_i}}$ \\
\hline
$S_i$ & Coverage area of ${\mathrm{CR_i}}$ \\
\hline
$l_i$ & Radius of ${\mathrm{CR_i}}$\\
\hline
$P_i$ & Number of users in ${\mathrm{CR_i}}$ \\
\hline
${N_S}\left( i \right)$ & Number of SBSs in ${\mathrm{CR_i}}$ \\
\hline
$m_i$ & Attraction exponent of ${\mathrm{HS_i}}$\\
\hline
$T_{ij}$ & Number of moving users from $\mathrm{HS_i}$ to ${\mathrm{HS}_j}$\\
\hline
${U_{\mathrm{HS}_i}}$ & Typical user of ${\mathrm{HS_i}}$ \\
\hline
${S_{\mathrm{HS}_i}}$ & Typical SBS of ${\mathrm{HS_i}}$ \\
\hline
\end{tabular}
\end{table}

In this paper, the R\&A model is used to analyze the distribution of mobile users in a two dimension plane $\mathbb{R}^2$. Let $\mathbb{A}_T$  be a region with area ${S_A}$  on the plane $\mathbb{R}^2$. The places with large number of users are denoted as hot spots in $\mathbb{A}_T$. The set of hot spots is denoted by $\Upsilon$, the cardinality of $\Upsilon$ is ${N_p}$ and elements of $\Upsilon$ are denoted by $\left\{ {\mathrm{\mathrm{HS}_i}|1 \le i \le {N_p}} \right\}$. The system $\mathrm{QN}$ is defined as a set formed by all users, hot spots and SBSs in $\mathbb{A}_T$. The coverage region of the hot spot $\mathrm{HS_i}$ is denoted by ${\mathrm{CR}_i}$ which is assumed to be a circle with a radius ${l_i}$. Thus, the area of ${\mathrm{CR}_i}$ is ${S_i} = \pi l_i^2$. According to the definition of the R\&A model, users can commute with all other hot spots in $\mathbb{A}_T$. The probability that one user moves from $\mathrm{HS_i}$ to ${\mathrm{HS}_j}$ is denoted as ${r_{ij}}$ and expressed as follow \cite{Simini12}.

\[{r_{ij}} = \frac{{{m_i}{m_j}}}{{\left( {{m_i} + {s_{ij}}} \right)\left( {{m_i} + {m_j} + {s_{ij}}} \right)}},\tag{1}\]
where ${m_i}$ and ${m_j}$ are attraction exponents (AEs) of $\mathrm{HS_i}$ and ${\mathrm{HS}_j}$, respectively. Based on the definition in the supplementary material of \cite{Simini12}, AEs are configured as fixed values determined by several parameters like the number of available jobs, average salary and consumption level of the corresponding hot spot. ${s_{ij}}$ denotes the total value of AEs (excluding AEs of $\mathrm{HS_i}$ and ${\mathrm{HS}_j}$) in the circle centered at $\mathrm{HS_i}$ with radius ${l_{ij}}$, where ${l_{ij}}$ is the distance between the centers of $\mathrm{HS_i}$ and ${\mathrm{HS}_j}$.

\begin{figure*}
\vspace{0.1in}
\centering
\includegraphics[width=10cm,draft=false]{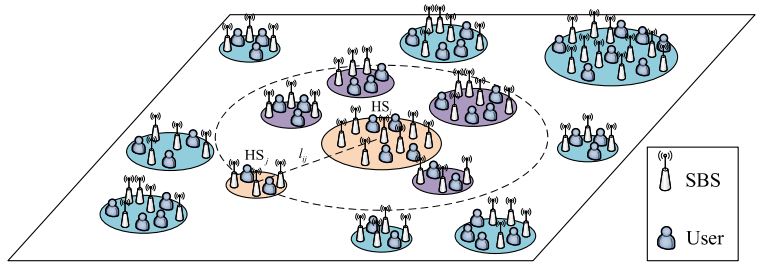}
\caption{\small System model. Coverage areas of different hot spots are shown as circles with different radius, ${s_{ij}}$  is the total number of AEs in circles marked as the purple regions.}
\end{figure*}

The total number of users commuting from $\mathrm{HS_i}$ to ${\mathrm{HS}_j}$ during $\Delta t$ is denoted by ${T_{ij}}$. Then, ${T_{ij}}$ is a binomial random variable by assuming moving characteristics of different users inside the QN system are independent from each other. The expectation of ${T_{ij}}$ is

\[\mathbb{E}\left( {{T_{ij}}} \right) = \zeta {P_i}\frac{{{m_i}{m_j}}}{{\left( {{m_i} + {s_{ij}}} \right)\left( {{m_i} + {m_j} + {s_{ij}}} \right)}},\tag{2}\]
where $\mathbb{E}\left( \cdot \right)$ is the expectation operation. The ratio of moving users in the total population is configured as a fixed value denoted by $\zeta$.

\subsection{Queueing Network Model}

 Based on the definition in \cite{Simini12}, the original R\&A model is only used for the theoretical analysis in the spatial domain. Thus, to evaluate the stationary number of users in the coverage region of each hot spot in the $\mathrm{QN}$ system, the queueing network model is adopted in this paper to extend the R\&A model to the temporal domain. We assume that all users in the coverage region of a hot spot are stayed in a queue so that the number of users in the coverage region is equal to the length of the corresponding queue. Furthermore, users in the queue are assumed to be served by a server in the corresponding hot spot, and a user will leave the hot spot after the user is served. Thus, the arrival rate and serving rate of a queue are evaluated by the expected number of users that moving in and out of the hot spot respectively during a given time slot $\Delta t$.

The Jackson network (JN) is a basic type of queueing networks. Based on the definition of JN \cite{Jackson54}, users outside of a queueing network system may move into a hot spot contained in the queueing network system. Fig. 2 is a Jackson network system (JNS) formed by two queues.

\begin{figure}[H]
\vspace{0.1in}
\centering
\includegraphics[width=8cm,draft=false]{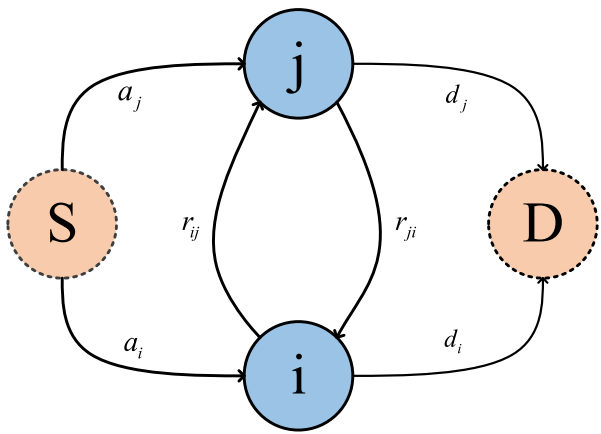}
\caption{\small An example of JNS that contains two queues $i$  and $j$. Users outside of JNS are denoted by $S$, and users exiting from JNS are denoted by $D$. The probability that a user transfers from $S$  to $i$  and $j$  is   ${a_i}$ and ${a_j}$.  The probability that a user leaves the JNS from $i$ ($j$)  is ${d_i}$ (${d_j}$). The probabilities that a user moves between  $i$ and $j$ are denoted by ${r_{ij}}$  and ${r_{ji}}$, respectively.}
\end{figure}

In this paper, the number of users in ${\mathrm{CR}_i}$ at time $t$ is denoted as ${P_i}$. Based on the definition of JN in \cite{Jackson54}, ${P_i}$ is independent from the length of other queues in the $\mathrm{QN}$ system. The serving rate of the queue in $\mathrm{HS_i}$ is denoted as ${\mu _i}$.  In the  $\mathrm{QN}$ system, the stationary value of ${P_i}$ exists if the condition ${\rho _i} = \frac{{{\lambda _i}}}{{{\mu _i}}} < 1$ is satisfied. The $\mathrm{QN}$ system keeps a stationary state when lengths of all queues in the $\mathrm{QN}$ system remain stationary.

\subsection{Interference Model}
In this paper, all SBSs and users of $\mathrm{HS_i}$ are assumed to be randomly and uniformly distributed in ${\mathrm{CR}_i}$. Furthermore, an typical user in ${\mathrm{CR}_i}$ is donated by ${U_{\mathrm{HS}_i}}$  and an typical SBS in ${\mathrm{CR}_i}$ is denoted as ${S_{\mathrm{HS}_i}}$. The distance between ${U_{\mathrm{HS}_i}}$ and ${S_{\mathrm{HS}_i}}$ is denoted by ${l_{mc}}$. Based on the result in \cite{Demestichas76}, we have

\[{f_{{l_{mc}}}}\left( x \right) = \frac{{4x}}{{\pi l_i^2}}\left( {\arccos \frac{x}{{2{l_i}}} - \frac{x}{{2{l_i}}}\sqrt {1 - \frac{{{x^2}}}{{4l_i^2}}} } \right),\tag{3}\]
where ${f_{{l_{mc}}}}\left( x \right)$ is the probability density function (PDF) of the random variable ${l_{mc}}$. In this paper, the function ${f_X}\left( \cdot \right)$ is used to denote the PDF of a random variable $X$.

The orthogonal frequency division multiplexing (OFDM) technology is assumed to be adopted by all SBSs in the $\mathrm{QN}$ system. In this paper, SBSs that transmit useful signals to users are called the associated SBSs while SBSs that generate interference are called interfering SBSs. Users are assumed to receive downlink interference only from SBSs within the same hot spot with them. The small scale fading of all channels are assumed to be governed by independent and identically distributed ({\em{i.i.d.}}) Rayleigh distributions. By ignoring the shadowing, the coverage probability of ${S_{\mathrm{HS}_i}}$  to ${U_{\mathrm{HS}_i}}$ is expressed by

\[{P_{\mathrm{cover}}} = \Pr \left( {\frac{{{p_t}{G_s}R_s^{ - {\alpha _p}}}}{{{\sigma ^2} + \sum\limits_{i = 0}^{\overline {{N_{\mathrm{int}}}} } {{p_t}{G_i}R_i^{ - {\alpha _p}}} }} \ge {\gamma _0}} \right),\tag{4}\]
where $\Pr \left( \cdot \right)$ is the probability corresponding to the expression in parentheses, ${\gamma _0}$ is the threshold of the received signal to interference plus noise ratio (SINR) at user devices, ${p_t}$ is the transmit power consumption of a SBS, ${\sigma ^2}$ is the power of the additive white Gaussian noise (AWGN), ${\alpha _p}$ is the path loss exponent, $\overline {{N_{{\mathop{\rm int}} }}}$ is the average number of interfering SBSs, ${R_s}$ is the distance between a user and an associated SBS, ${R_i}$ is the distance between a user and an interfering SBS, ${G_s}$ and ${G_i}$ are the small scale fading experienced by the desired link and interfering links, respectively. Based on the assumption that small scale fading of all channels follows {\em{i.i.d.}} Rayleigh distributions, ${G_s}$ and ${G_i}$ are exponential distributed random variables with expectations $\eta $.

The event that a user is covered by a SBS is assumed to be independent of the event that the user is covered by another SBS. Thus, the number of SBSs that cover the same user is a binomial distributed random variable. The expectation of the binomial distributed random variable is

\[\mathbb{E}\left( {{N_{\mathrm{cover}}}} \right) = \overline {{N_{\mathrm{cover}}}}  = {N_S}\left( i \right){P_{\mathrm{cover}}}.\tag{5}\]

All SBSs that can cover a same user are configured to form a virtual cell cluster (VC). SBSs in a VC are assumed to be able to transmit signals to the served user simultaneously without causing interference \cite{Ge16}. Thus, the number of interfering SBSs is

\[\overline {{N_{\mathrm{int}}}}  = {N_{\mathrm{SBS}}} - \overline {{N_{\mathrm{cover}}}}  = {N_S}\left( i \right)\left( {1 - {P_{\mathrm{cover}}}} \right).\tag{6}\]

By substituting (6) into (4), the coverage probability is further derived by
{\small{
\[{P_{\mathrm{cover}}} = \Pr \left( {\frac{{{p_t}{G_s}R_s^{ - {\alpha _p}}}}{{{\sigma ^2} + \sum\nolimits_{i = 0}^{{N_S}\left( i \right)\left( {1 - {P_{\mathrm{cover}}}} \right)} {{p_t}{G_i}R_i^{ - {\alpha _p}}} }} \ge {\gamma _0}} \right).\tag{7}\]
}}

\subsection{Channel Access Model}

 A channel is called an available channel if the received SINR at the user using this channel is larger than a given threshold ${\gamma _0}$. A user will be blocked when all available channels of the associated SBS are occupied.  Thus, a two-dimensional Markov chain is utilized to analyze channel access processes of SBSs in the $\mathrm{QN}$ system \cite{Ge15}. Two states of the two-dimensional Markov chain are denoted by $\left( {{v_m},{v_n}} \right)$, where ${v_m}$ denotes the number of occupied channels in a SBS and ${v_n}$ is the number of available channels in a SBS (${v_m} \le {v_n}$), respectively.

The probability that a user has a downlink communication request to a SBS is denoted by ${p_s}$. Communication requests of different users are assumed to be {\em{i.i.d.}}. Thus, the number of users with communication requests follows a binomial distribution with expectation $\overline {{P_i}} \cdot {p_s}$, where $\overline {{P_i}} $ is the stationary number of users in ${\mathrm{CR}_i}$. Since $\overline {{P_i}}$ is usually a large value, the binomial distribution can be approximated by a Poisson distribution with the same expectation $\overline {{P_i}} \cdot {p_s}$. The call arriving process of ${S_{\mathrm{HS}_i}}$ is denoted by ${\mathrm{AP}}$ and the number of calls arriving at ${S_{\mathrm{HS}_i}}$ before time $t$ is denoted by ${\mathrm{AP}}\left( t \right)$. Thus, we have the following equation
{\footnotesize{
\[{\rm{Pr}}\left\{ {{\mathrm{AP}}\left( {t + \Delta t} \right) - {\mathrm{AP}}\left( t \right) = n} \right\} = {e^{ - \frac{{\overline {{P_i}} \cdot {p_s}\Delta t}}{{{N_S}\left( i \right)}}}}\frac{{\overline {{P_i}} \cdot {p_s}{{\left( {\Delta t} \right)}^n}}}{{n!{N_S}\left( i \right)}}.\tag{8}\]
}}

The call arriving process of ${S_{\mathrm{HS}_i}}$ is supposed to be a stochastic process with independent and stationary increments. Thus, the stochastic process $\mathrm{AP}$ is a Poisson process. The value of the serving duration of ${S_{\mathrm{HS}_i}}$ to ${U_{\mathrm{HS}_i}}$  is assumed to be an exponential distributed random variable with expectation ${\mu _s}$. The corresponding transition diagram of the two-dimensional Markov chain is shown as follows.

\begin{figure*}
\vspace{0.1in}
\centering
\includegraphics[width=10cm,draft=false]{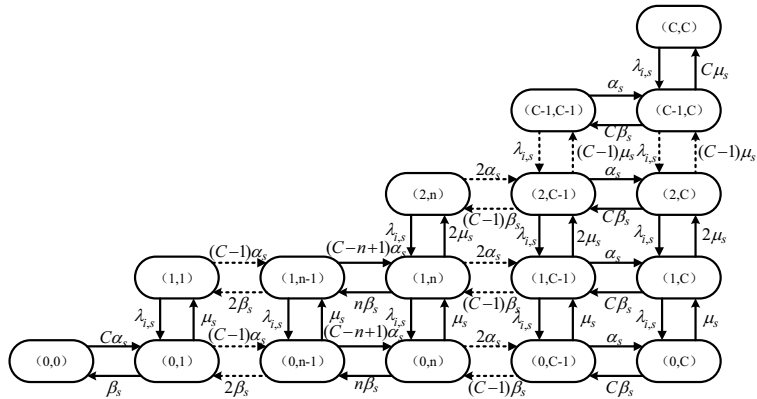}
\caption{\small Transition diagram of the two-dimensional Markov chain.}
\end{figure*}

The transition diagram of the two-dimensional Markov chain is explained as follows.

\begin{enumerate}
\item $\left( {{v_m},{v_n}} \right) \to \left( {{v_m} + 1,{v_n}} \right)$: A user has been successfully allocated an available idle channel of ${S_{\mathrm{HS}_i}}$, then ${v_m} + 1$.
\item $\left( {{v_m},{v_n}} \right) \to \left( {{v_m},{v_n} + 1} \right)$: An unavailable channel of ${S_{\mathrm{HS}_i}}$  becomes available due to the time-varying interference, thus ${v_n} + 1$.
\item $\left( {{v_m},{v_n}} \right) \to \left( {{v_m} - 1,{v_n}} \right)$: A user has been fully served and releases the occupied channel of ${S_{\mathrm{HS}_i}}$, then  ${v_m} - 1$.
\item $\left( {{v_m},{v_n}} \right) \to \left( {{v_m},{v_n} - 1} \right)$: An available channel of ${S_{\mathrm{HS}_i}}$  becomes unavailable due to the time-varying interference, thus ${v_n} - 1$.
\end{enumerate}

The maximum number of available channels offered by ${S_{\mathrm{HS}_i}}$  is denoted as $C$. The transition rate for an unavailable channel becoming available is denoted by  ${\alpha _s}$, and the transition rate for an available channel becoming unavailable is denoted by ${\beta _s}$.  The call arriving rate ${\lambda _{i,s}}$ is

\[{\lambda _{i,s}} = \frac{{\overline {{P_i}} \cdot {p_s} \cdot \overline {{N_{\mathrm{cover}}}} }}{{{N_S}\left( i \right)}}.\tag{9}\]

Based on the Kolmogorov criteria \cite{Ge15}, the two-dimensional Markov chain in Fig. 3 is reversible and the stationary state distribution of the two-dimensional Markov chain exists.

\section{Coverage Probability, Blocking Probability and Throughput}
\subsection{Coverage Probability}

The coverage probability in (7) can be further evaluated as (10).

{\footnotesize{
\[\begin{array}{l}
{P_{{\rm{cover}}}} = \Pr \left( {\frac{{{p_t}{G_s}R_s^{ - {\alpha _p}}}}{{{\sigma ^2} + \sum\nolimits_{i = 0}^{{N_S}\left( i \right)\left( {1 - {P_{\mathrm{cover}}}} \right)} {{p_t}{G_i}R_i^{ - {\alpha _p}}} }} \ge {\gamma _0}} \right)\\
\quad \quad \; = \Pr \left( {\frac{{{G_s}}}{{\frac{1}{{{\rho _0}}} + \frac{{\sum\nolimits_{i = 0}^{{N_S}\left( i \right)\left( {1 - {P_{\mathrm{cover}}}} \right)} {{G_i}R_i^{ - {\alpha _p}}} }}{{R_s^{ - {\alpha _p}}}}}} \ge {\gamma _0}} \right)\\
\quad \quad \; = \Pr \left( {\frac{1}{{{\rho _0}{G_s}}} + \frac{{\sum\nolimits_{i = 0}^{{N_S}\left( i \right)\left( {1 - {P_{\mathrm{cover}}}} \right)} {{G_i}R_i^{ - {\alpha _p}}} }}{{{G_s}R_s^{ - {\alpha _p}}}} \le \gamma _0^{ - 1}} \right),
\end{array}\normalsize{\tag{10}}\]
}}
where ${\rho _0}$  is the received signal to noise ratio (SNR) of ${U_{\mathrm{HS}_i}}$. The expression of  ${\rho _0}$  is

\[{\rho _0} = \frac{{{p_t}R_s^{ - {\alpha _p}}}}{{{\sigma ^2}}}.\tag{11}\]

Based on the result in \cite{Guo14}, ${\rho _0}$  is configured as a determined variable in this paper.

Based on (10) and (11), the coverage probability $P_{\mathrm{cover}}$ is finally derived as

\begin{figure*}[!t]
{\footnotesize{
\[\begin{array}{l}
{P_{\mathrm{cover}}} = {2^{ - {B_e}}}{\gamma _0}\exp \left( {\frac{{{A_e}}}{2}} \right)\sum\limits_{{b_e} = 0}^{{B_e}} {\left( \begin{array}{l}
{B_e}\\
{b_e}
\end{array} \right)} \sum\limits_{{c_e} = 0}^{{C_e} + {b_e}} {\frac{{{{\left( { - 1} \right)}^{{c_e}}}}}{{{D_e}}}}\times\\
\quad \quad \;\;{\kern 1pt} \,{\mathop{\rm Re}\limits} \left( {\left( {\int\limits_0^\infty  {\left( {\eta \exp \left( { - \frac{\hbar }{{{\rho _0}{G_s}}} - \eta {G_s}} \right)} \right.\left( {\int\limits_0^\infty  {\int\limits_0^{2{l_i}} {\int\limits_0^{2{l_i}} {\exp \left( {\frac{{ - \hbar {G_i}R_i^{ - {\alpha _p}}}}{{{G_s}R_s^{ - {\alpha _p}}}}} \right)\left( {\frac{{4{R_s}}}{{\pi l_i^2}}\left( {\arccos \frac{{{R_s}}}{{2{l_i}}} - \frac{{{R_s}}}{{2{l_i}}}\sqrt {1 - \frac{{{R_s}^2}}{{4l_i^2}}} } \right)} \right)} } } } \right.} } \right.} \right.\times \\
\quad \quad \;\;{\kern 1pt} \quad \left. {\left. {\left. {{{\left. {\left( {\frac{{4{R_i}}}{{\pi l_i^2}}\left( {\arccos \frac{{{R_i}}}{{2{l_i}}} - \frac{{{R_i}}}{{2{l_i}}}\sqrt {1 - \frac{{{R_i}^2}}{{4l_i^2}}} } \right)} \right)\left( {\eta \exp \left( { - \eta {G_i}} \right)} \right)d{G_i}d{R_s}d{R_i}} \right)}^{\overline {{N_{\mathrm{int}}}} }}} \right)d{G_s}} \right)/\hbar } \right).
\end{array}\normalsize{\tag{12}}\]
}}
\end{figure*}

The corresponding derivation and explanation is shown in the appendix A.

\subsection{Blocking Probability}

By the definition of JN in \cite{Jackson54}, a $\mathrm{QN}$ system should meet the following three necessary conditions for being a JN:

\begin{enumerate}
\item User arriving processes in all hot spots are independent Poisson processes in the $\mathrm{QN}$  system
\item Users outside of the $\mathrm{QN}$ system can move into the $\mathrm{QN}$ system.
\item Users who are fully served in the $\mathrm{QN}$  system can either move to another hot spot in the $\mathrm{QN}$  system or leave the $\mathrm{QN}$ system.
\end{enumerate}

Denote the total number of users moving from outside of $\mathbb{A}_T$  into ${\mathrm{CR}_i}$ during a time period $\Delta t$ as ${e_i}$. By assuming ${e_i}$ as a random variable with expectation $a_i$, we get the following lemmas.

\textbf{\em Lemma 1}: For a large $P_i$, the binomial distributed random variable ${T_{ij}}$  can be approximated by a Poisson distributed random variable with the expectation being ${\mathbb{E}\left( {{T_{ij}}} \right)}$.

The number of users in a hot spot is modeled by a queueing network model in the temporal domain. Based on the definition of JN users who have been fully served in a hot spot either choose to be added into another queue or to leave the queueing network \cite{Jackson54}. Based on the above model, we get the following Lemma 2.

\textbf{\em Lemma 2}: Based on Lemma 1 and the assumption that ${e_i}$  is a Poisson distributed random variable, the number of users moving into ${\mathrm{CR}_i}$ during $\Delta t$  is a Poisson distributed random variable. The expectation of the number of users moving into ${\mathrm{CR}_i}$ during $\Delta t$ is

\[{\lambda _i} = {a_i} + \sum\limits_{j \in \left( {1,{N_p}} \right),j \ne i} {\mathbb{E}\left( {{T_{ji}}} \right)}.\tag{13}\]

\textbf{\em Lemma 3}: Assuming numbers of users moving into ${\mathrm{CR}_i}$ during each disjoint $\Delta t$  in the temporal domain are {\em{i.i.d.}}. Thus, the user arriving process of $\mathrm{HS_i}$ is a stochastic process with independent and stationary increments.

By combining Lemma 1, Lemma 2 and Lemma 3, the user arriving process of $\mathrm{HS_i}$  is proved to be a Poisson process. Then, we get the following theorem.

\textbf{\em Theorem 1}: Assuming users' moving actions among different hot spots are governed by a R\&A model. By denoting the number of users in each hot spot as the size of the corresponding queue, the $\mathrm{QN}$  system can be modeled by a Jackson network.

The transferring matrix of the $\mathrm{QN}$  system is denoted by ${\bf{R}}$. The element of the matrix ${\bf{R}}$  located at the row $i$  and the line $j$ is denoted as ${\left[ {\bf{R}} \right]_{ij}}$, which is the probability that a user moves from $\mathrm{HS_i}$ to ${\mathrm{HS}_j}$. Thus, ${\left[ {\bf{R}} \right]_{ij}}$  is expressed by
{\small{
\[{\left[ {\bf{R}} \right]_{ij}} = \left\{ \begin{array}{l}
\frac{{{m_i}{m_j}}}{{\left( {{m_i} + {s_{ij}}} \right)\left( {{m_i} + {m_j} + {s_{ij}}} \right)}} \quad 0 < i \ne j \le {N_p}\\
\quad \quad \quad \quad 0\quad \quad \quad \quad \quad \;\;\, 0 < i = j \le {N_p}
\end{array} \right..\normalsize{\tag{14}}\]
}}

The probability that a user leaves the $\mathrm{QN}$  system from $\mathrm{HS_i}$ is

\[{d_i} = 1 - \sum\limits_{j = 1}^{{N_p}} {{{\left[ {\bf{R}} \right]}_{ij}}}.\tag{15}\]

When the $\mathrm{QN}$  system is in a stationary state the number of users leave the $\mathrm{QN}$ system from $\mathrm{HS_i}$ during $\Delta t$ is denoted as $\overline {{D_i}}$ with expectation as follows.

\[\mathbb{E}\left( {\overline {{D_i}} } \right) = \zeta \overline {{P_i}},\tag{16}\]
where $\overline {{P_i}}$ denotes the stationary number of users in ${\mathrm{CR}_i}$. Substitute (14) and (16) into (2), the expectation of the number of users moving into ${\mathrm{CR}_i}$ during $\Delta t$ is

\[{\lambda _i} = {a_i} + \sum\limits_{j \in \left( {1,{N_p}} \right),j \ne i} {{\lambda _j}{r_{ji}}}.\tag{17}\]

To keep the stability of the total number of users in the $\mathrm{QN}$ system, ${a_i}$ is configured as

\[{a_i} = \zeta \frac{{\sum\limits_{i = 1}^{{N_p}} {{m_i}{d_i}} }}{{{N_p}}} = \zeta \frac{{\sum\limits_{i = 1}^{{N_p}} {{m_i}\left( {1 - \sum\limits_{j = 1}^{{N_p}} {{{\left[ {\bf{R}} \right]}_{ij}}} } \right)} }}{{{N_p}}}.\tag{18}\]

By denoting ${\bf{a}} = {\left( {{a_1},{a_2}, \ldots ,{a_{{N_p}}}} \right)_{1 \times {N_p}}}$  and ${\bm{\lambda }} = {\left( {{\lambda _1},{\lambda _2}, \ldots ,{\lambda _{{N_p}}}} \right)_{1 \times {N_p}}}$, (28) is further derived by a matrix form as

\[\bm{\lambda}  = \bf{a}{\left( {{\bf{I - R}}} \right)^{ - 1}},\tag{19}\]
where ${\bf{I}}$  is an identical matrix. Based on (19), the user arriving rate of each hot spot can be obtained.

The stationary serving rate of the queue in $\mathrm{HS_i}$  is denoted as ${\mu _i}$. By substituting ${\lambda _i}$  into  ${\rho _i} = \frac{{{\lambda _i}}}{{{\mu _i}}} = \frac{{{\lambda _i}}}{{\zeta \overline {{P_i}} }}$, the number of users in the stable status in ${\mathrm{CR}_i}$ is

\[\overline {{P_i}}  = \frac{{{\rho _i}}}{{1 - {\rho _i}}} = \frac{{{\lambda _i}}}{{\zeta \overline {{P_i}}  - {\lambda _i}}}.\tag{20}\]

By solving the implicit function in (11),  $\overline {{P_i}}$ is further derived by

\[\overline {{P_i}}  = \sqrt {{{\left( {\frac{{{\lambda _i}}}{{2\zeta }}} \right)}^2} + \frac{{{\lambda _i}}}{\zeta }}  + \frac{{{\lambda _i}}}{{2\zeta }}.\tag{21}\]

The stationary distribution of the state $\left( {{v_m},{v_n}} \right)$  in the two-dimensional Markov chain in Fig. 3 is denoted by $\pi \left( {{v_m},{v_n}} \right)$. Based on the result in \cite{Ge15}, $\pi \left( {{v_m},{v_n}} \right)$ is expressed by (22)
\begin{figure*}[!t]
\[\pi \left( {{v_m},{v_n}} \right) = \left\{ \begin{array}{l}
\frac{1}{\chi }{\left( {\frac{{{\lambda _{i,s}}}}{{{\mu _s}}}} \right)^{{v_m}}}\frac{1}{{{v_m}!}}\left(\small{ \begin{array}{l}
C\\
{v_n}
\end{array}} \right){\left( {\frac{{{\alpha _s}}}{{{\beta _s}}}} \right)^{{v_n}}}\\
\chi  = \sum\limits_{{v_m} \le {v_n} \le C} {{{\left( {\frac{{{\lambda _{i,s}}}}{{{\mu _s}}}} \right)}^{{v_m}}}\frac{1}{{{v_m}!}}\left(\small{ \begin{array}{l}
C\\
{v_n}
\end{array}} \right){{\left( {\frac{{{\alpha _s}}}{{{\beta _s}}}} \right)}^{{v_n}}}}
\end{array} \right.,\tag{22}\]
\end{figure*}
where $\small{\left( \begin{array}{l}
C\\
{v_n}
\end{array} \right)}$  is a binomial coefficient denoting the number of ways to pick $v_n$ unordered outcomes from $C$ possibilities. By the Gilbert-Elliott model \cite{Ge15}, we have

\[\frac{{{\alpha _s}}}{{{\beta _s}}} = \frac{{1 - {\varepsilon _s}}}{{{\varepsilon _s}}},\tag{23}\]
where ${\varepsilon _s}$ is expressed as (35) and ${p_{\mathrm{oc}}}$  is the probability that a channel of ${S_{\mathrm{HS}_i}}$  is occupied.
{\scriptsize{
\[{\varepsilon _s} = \sum\limits_{{\Delta _s} = 0}^{\overline {{N_{\mathrm{int}}}} } {\left( {1 - {P_{\mathrm{cover}}}} \right)} \left( \begin{array}{l}
\overline {{N_{\mathrm{int}}}} \\
\;{\Delta _s}
\end{array} \right){\left( {{p_{\mathrm{oc}}}} \right)^{{\Delta _s}}}{\left( {1 - {p_{\mathrm{oc}}}} \right)^{\overline {{N_{\mathrm{int}}}}  - {\Delta _s}}},\normalsize{\tag{24}}\]
}}

By assuming ${U_{\mathrm{HS}_i}}$  accesses each channel with equal probability, the probability that a channel of ${S_{\mathrm{HS}_i}}$  is occupied is

\[{p_{\mathrm{oc}}} = \sum\limits_{{v_m} = 0}^{{v_n}} {\sum\limits_{{v_n} = 0}^C {\frac{{{v_m}}}{C}} } \pi \left( {{v_m},{v_n}} \right).\tag{25}\]

By substituting (23) (24) (25) into (22), the stationary probability  $\pi \left( {{v_m},{v_n}} \right)$ of the state   $\left( {{v_m},{v_n}} \right)$ is obtained. The call blocking probability of ${U_{\mathrm{HS}_i}}$  is

\[{P_B} = \sum\limits_{{v_m} = {v_n} \le C} {\pi \left( {{v_m},{v_n}} \right)}.\tag{26}\]

By (22) and (26), the probability that ${U_{\mathrm{HS}_i}}$  is served by ${S_{\mathrm{HS}_i}}$ is derived as

{\small{
\[\begin{array}{l}
{P_{{\rm{serve}}}} = 1 - {P_B}\\
{\kern 1pt} {\kern 1pt}  = 1 - \sum\limits_{{v_m} = {v_n} \le C} {\frac{1}{\chi }} {\left( {\frac{{{\lambda _{i,s}}}}{{{\mu _s}}}} \right)^{{v_m}}}\frac{1}{{{v_m}!}}\left( \begin{array}{l}
C\\
{v_n}
\end{array} \right){\left( {\frac{{{\alpha _s}}}{{{\beta _s}}}} \right)^{{v_n}}}
\end{array}.\normalsize{\tag{27}}\]
}}

The probability that all channels of ${S_{\mathrm{HS}_i}}$  are idle is ${P_{\mathrm{idle}}}$, given by

\[{P_{\mathrm{idle}}} = \sum\limits_{{v_n} = 0}^C {\pi \left( {0,{v_n}} \right)}.\tag{28}\]

\subsection{Network Throughput}

Based on the transition diagram of the two-dimensional Markov Chain in Fig. 3, the average number of occupied channels at ${S_{\mathrm{HS}_i}}$  is

\[\overline {{N_{\mathrm{oc}}}}  = \sum\limits_{{v_m} = 0}^{{v_n}} {\sum\limits_{{v_n} = 0}^C {{v_m}} } \pi \left( {{v_m},{v_n}} \right).\tag{29}\]

We assume that the transmission rate is assumed to be equal to the channel capacity when the channel is occupied. Thus, the stationary throughput of ${S_{\mathrm{HS}_i}}$ is
\[\begin{array}{l}
{T_s} = \overline {{N_{{\rm{oc}}}}} \cdot \left( {{C_{ca}}} \right)\\
\;\;\;{\kern 1pt}  = \left( {{C_{ca}}} \right)\sum\limits_{{v_m} = 0}^{{v_n}} {\sum\limits_{{v_n} = 0}^C {{v_m}} } \pi \left( {{v_m},{v_n}} \right)\quad ,
\end{array} \tag{30}\]

By assuming the throughputs of all SBSs in the $\mathrm{QN}$ system are {\em{i.i.d.}}, the throughput of the small cell network restricted in ${\mathrm{CR}_i}$ is (31).
\[\begin{array}{l}
{T_{\mathrm{HS}_i}} = {N_S}\left( i \right)\left( {\int\limits_{x = {\gamma _0}}^\infty  {{{\log }_2}\left( {1 + x} \right){f_{\mathchar'26\mkern-10mu\lambda} }\left( x \right)dx} } \right) \times \\
\;\;\;\;\;\;\;\;\;{\kern 1pt}\sum\limits_{{v_m} = 0}^{{v_n}} {\sum\limits_{{v_n} = 0}^C {{v_m}} } \pi \left( {{v_m},{v_n}} \right).
\end{array} \tag{31}\]
The throughput of the small cell network in the $\mathrm{QN}$ system is (32).
\[\begin{array}{l}
{T_{\mathrm{QN}}} = \sum\limits_{i = 1}^{{N_p}} {{N_S}\left( i \right)\left( {\int\limits_{x = {\gamma _0}}^\infty  {{{\log }_2}\left( {1 + x} \right){f_{\mathchar'26\mkern-10mu\lambda} }\left( x \right)dx} } \right)} \times \\
\;\;\;\;\;\;\;\;\;{\kern 1pt}\sum\limits_{{v_m} = 0}^{{v_n}} {\sum\limits_{{v_n} = 0}^C {{v_m}} } \pi \left( {{v_m},{v_n}} \right).
\end{array} \normalsize{\tag{32}}\]
\subsection{Energy Efficiency}

The total power consumption ${p_{\mathrm{to}}}$ of a SBS ${s_{\mathrm{ori}}}$  is \cite{Imran11}

\[{p_{\mathrm{to}}} = {N_{\mathrm{oc}}}\frac{{\frac{{{p_t}}}{{{\eta _{\mathrm{pa}}}}} + {p_\mathrm{{rf}}} + {p_{\mathrm{bb}}}}}{{\left( {1 - {\sigma _{\mathrm{dc}}}} \right)\left( {1 - {\sigma _{\mathrm{ms}}}} \right)}} + {p_{\mathrm{st}}},\tag{33}\]
where ${p_t}$  is the transmit power consumption of ${s_{\mathrm{ori}}}$, ${\eta _{\mathrm{pa}}}$ is the efficiency coefficient of the power amplify module,  ${p_\mathrm{{rf}}}$  is the power consumption of the radio frequency module, ${p_{\mathrm{bb}}}$ is the power consumption of the base band module, ${\sigma _{\mathrm{dc}}}$  is the loss coefficient of the digital control module, ${\sigma _{\mathrm{ms}}}$  is the power supply loss coefficient,  ${p_{\mathrm{st}}}$ is the constant power consumption which is independent from the traffic load of ${s_{\mathrm{ori}}}$, and ${N_{\mathrm{oc}}}$  is the number of occupied channels at ${s_{\mathrm{ori}}}$, respectively. Based on the configuration in \cite{Imran11}, the values of ${\sigma _{\mathrm{ms}}}$ and ${\sigma _{\mathrm{dc}}}$ are smaller than 1.

We assume that all SBSs are able to be shut down and turned on instantaneously without delay. Thus, the energy consumption of ${S_{\mathrm{HS}_i}}$ is

\[\begin{array}{l}
{E_{to}} = \left( {\sum\limits_{{v_m} = 1}^{{v_n}} {\sum\limits_{{v_n} = 1}^C {{v_m}} } \pi \left( {{v_m},{v_n}} \right)\frac{{\frac{{{p_t}}}{{{\eta _{\mathrm{pa}}}}} + {p_\mathrm{{rf}}} + {p_{\mathrm{bb}}}}}{{\left( {1 - {\sigma _{\mathrm{dc}}}} \right)\left( {1 - {\sigma _{\mathrm{ms}}}} \right)}} + {p_{\mathrm{st}}}} \right)\times \\
\;\;\;\;\;\;\;\;\;\;\;\left( {1 - {P_{\mathrm{idle}}}} \right)\cdot {t_{to}},
\end{array}\normalsize{\tag{34}}\]
Where ${t_{to}}$ is the operation duration of ${S_{\mathrm{HS}_i}}$.

By assuming the energy consumptions of different SBSs to be {\em{i.i.d.}}, the energy consumed by the small cell network restricted in ${\mathrm{CR}_i}$ is
{\small{
\[\begin{array}{l}
{E_{\mathrm{HS}_i}} = \left( {\sum\limits_{{v_m} = 1}^{{v_n}} {\sum\limits_{{v_n} = 1}^C {{v_m}} } \pi \left( {{v_m},{v_n}} \right)\frac{{\frac{{{p_t}}}{{{\eta _{\mathrm{pa}}}}} + {p_\mathrm{{rf}}} + {p_{\mathrm{bb}}}}}{{\left( {1 - {\sigma _{\mathrm{dc}}}} \right)\left( {1 - {\sigma _{\mathrm{ms}}}} \right)}} + {p_{\mathrm{st}}}} \right)\times\\
\;\;\;\;\;\;\;\;\;\;\;\,{N_S}\left( i \right)\cdot\left( {1 - {P_{\mathrm{idle}}}} \right)\cdot{t_{to}}.
\end{array}\normalsize{\tag{35}}\]
}}

Thus, the network energy consumption of the small cell network in the $\mathrm{QN}$ system is

{\small{
\[\begin{array}{l}
{E_{\mathrm{QN}}} = \sum\limits_{i = 1}^{{N_p}}  \left( {\sum\limits_{{v_m} = 1}^{{v_n}} {\sum\limits_{{v_n} = 1}^C {{v_m}} } \pi \left( {{v_m},{v_n}} \right)\frac{{\frac{{{p_t}}}{{{\eta _{\mathrm{pa}}}}} + {p_\mathrm{{rf}}} + {p_{\mathrm{bb}}}}}{{\left( {1 - {\sigma _{\mathrm{dc}}}} \right)\left( {1 - {\sigma _{\mathrm{ms}}}} \right)}} + {p_{\mathrm{st}}}} \right)\times\\
\;\;\;\;\;\;\;\;\;\;\;\,{{N_S}\left( i \right)}\cdot\left( {1 - {P_{\mathrm{idle}}}} \right)\cdot{t_{to}}.
\end{array}\normalsize{\tag{36}} \]}}

By combining (30) and (34), the energy efficiency of ${S_{\mathrm{HS}_i}}$ will be (48).

\begin{figure*}[!t]
\[{EE_s} = \frac{{\left( {\int\limits_{x = {\gamma _0}}^\infty  {{{\log }_2}\left( {x + 1} \right){f_{{\mathchar'26\mkern-10mu\lambda}} }\left( x \right)dx} } \right)\sum\limits_{{v_m} = 0}^{{v_n}} {\sum\limits_{{v_n} = 0}^C {{v_m}} } \pi \left( {{v_m},{v_n}} \right)}}{{\left( {\sum\limits_{{v_m} = 1}^{{v_n}} {\sum\limits_{{v_n} = 1}^C {{v_m}} } \pi \left( {{v_m},{v_n}} \right)\frac{{\frac{{{p_t}}}{{{\eta _{{\rm{pa}}}}}} + {p_{{\rm{rf}}}} + {p_{{\rm{bb}}}}}}{{\left( {1 - {\sigma _{{\rm{dc}}}}} \right)\left( {1 - {\sigma _{{\rm{ms}}}}} \right)}} + {p_{\mathrm{st}}}} \right)\left( {1 - {P_{\mathrm{idle}}}} \right)}}.\tag{48}\]
\end{figure*}

By combining (31) and (35), the network energy efficiency restricted in ${\mathrm{CR}_i}$ is (38).

\begin{figure*}[!t]
\[E{E_{\mathrm{HS}_i}} = \frac{{{N_S}\left( i \right)\left( {\int\limits_{x = {\gamma _0}}^\infty  {{{\log }_2}\left( {x + 1} \right){f_{{\mathchar'26\mkern-10mu\lambda}} }\left( x \right)dx} } \right)\sum\limits_{{v_m} = 0}^{{v_n}} {\sum\limits_{{v_n} = 0}^C {{v_m}} } \pi \left( {{v_m},{v_n}} \right)}}{{{N_S}\left( i \right)\left( {\sum\limits_{{v_m} = 1}^{{v_n}} {\sum\limits_{{v_n} = 1}^C {{v_m}} } \pi \left( {{v_m},{v_n}} \right)\frac{{\frac{{{p_t}}}{{{\eta _{\mathrm{pa}}}}} + {p_\mathrm{{rf}}} + {p_{\mathrm{bb}}}}}{{\left( {1 - {\sigma _{\mathrm{dc}}}} \right)\left( {1 - {\sigma _{\mathrm{ms}}}} \right)}} + {p_{\mathrm{st}}}} \right)\left( {1 - {P_{\mathrm{idle}}}} \right)}}.\tag{38}\]
\end{figure*}

Based on (32) and (36), the energy efficiency of the small cell network in the $\mathrm{QN}$  system is (39).

\begin{figure*}[!t]
\[E{E_{\mathrm{QN}}} = \frac{{\sum\limits_{i = 1}^{{N_p}} {{N_S}\left( i \right)\left( {\int\limits_{x = {\gamma _0}}^\infty  {{{\log }_2}\left( {x + 1} \right){f_{{\mathchar'26\mkern-10mu\lambda}} }\left( x \right)dx} } \right)} \sum\limits_{{v_m} = 0}^{{v_n}} {\sum\limits_{{v_n} = 0}^C {{v_m}} } \pi \left( {{v_m},{v_n}} \right)}}{{\sum\limits_{i = 1}^{{N_p}} {{N_S}\left( i \right)} \left( {\sum\limits_{{v_m} = 1}^{{v_n}} {\sum\limits_{{v_n} = 1}^C {{v_m}} } \pi \left( {{v_m},{v_n}} \right)\frac{{\frac{{{p_t}}}{{{\eta _{\mathrm{pa}}}}} + {p_\mathrm{{rf}}} + {p_{\mathrm{bb}}}}}{{\left( {1 - {\sigma _{\mathrm{dc}}}} \right)\left( {1 - {\sigma _{\mathrm{ms}}}} \right)}} + {p_{\mathrm{st}}}} \right)\left( {1 - {P_{\mathrm{idle}}}} \right)}}.\tag{39}\]
\end{figure*}

\section{Numerical and simulation results}

Based on our proposed 5G ultra-dense small cell network model in this paper, the numerical and simulation results are shown in this section. Some default parameters are configured as ${l_i} = 1\,{\rm{km}}$, $\zeta  = 0.1$, ${\alpha _p} = 4$, ${\rho _0} = 100$, $\eta  = 1$, ${p_s} = 0.01$, $C = 10$, ${p_t} = 1.6\,{\rm{W}}$, ${\eta _{\mathrm{pa}}} = 8$, ${p_\mathrm{{rf}}} = 0.7\,{\rm{W}}$, ${p_{\mathrm{bb}}} = 1.6\,{\rm{W}}$, ${p_{\mathrm{st}}} = 6.8\,{\rm{W}}$, ${\sigma _{\mathrm{dc}}} = 0.08$, ${\sigma _{\mathrm{ms}}} = 0.1$. Without loss of generality, the centers of hot spots are uniformly distributed in ${\mathbb{A}_T}$. AEs of hot spots are configured to be {\em{i.i.d.}} Poisson distributed with expectation 3000. The ratio of the stationary user density to the SBS density in ${\mathrm{CR}_i}$ is ${\theta _i} = \frac{{\overline {{P_i}} }}{{{v_n}\left( i \right)}}$.

\begin{figure}[H]
\vspace{0.1in}
\centering
\includegraphics[width=8cm, draft=false]{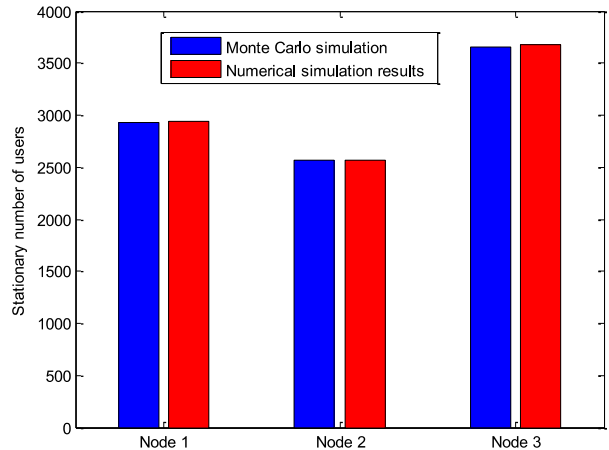}
\caption{\small  Comparison between theoretical results and simulation for the stationary number of users in a hot spot.}
\end{figure}

Comparison of the stationary number of users in a hot spot between numerical results and Monte Carlo simulation is shown in Fig. 4. Three hot spots are selected from  ${N_p} = 50$  hot spots to make this comparison. It is observed from the figure that gaps between Monte carlo simulation and numerical results are very small, indicating that using queueing network theory to extend the R\&A model to the temporal domain for the analysis of user density is practicable.

\begin{figure}[H]
\vspace{0.1in}
\centering
\includegraphics[width=8cm, draft=false]{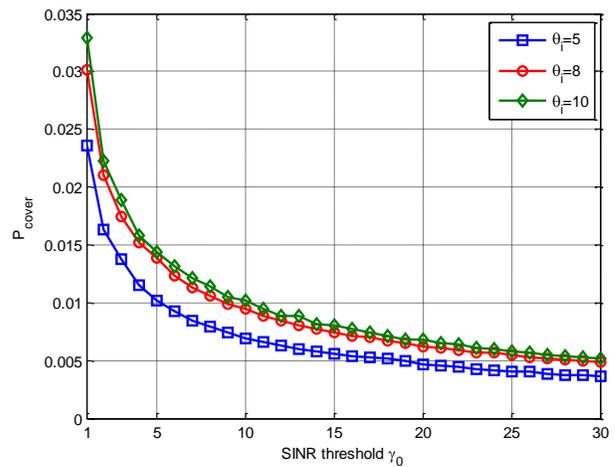}
\caption{\small  Coverage probability with respect to the SINR threshold ${\gamma _0}$  considering different ratios of the stationary user density to the SBS density ${\theta _i}$.}
\end{figure}

The impact of the SINR threshold ${\gamma _0}$  and the ratio ${\theta _i}$  on the coverage probability  ${P_{\mathrm{cover}}}$ is investigated in Fig. 5. Fig. 5 shows that when ${\gamma _0}$  is fixed the coverage probability ${P_{\mathrm{cover}}}$  decreases when increasing ${\theta _i}$. When  ${\theta _i}$ is fixed, the coverage probability ${P_{\mathrm{cover}}}$  decreases with the increase of ${\gamma _0}$. The observations is consistent with the results in \cite{Ge16,Ge15}.

\begin{figure}[H]
\vspace{0.1in}
\centering
\includegraphics[width=8cm, draft=false]{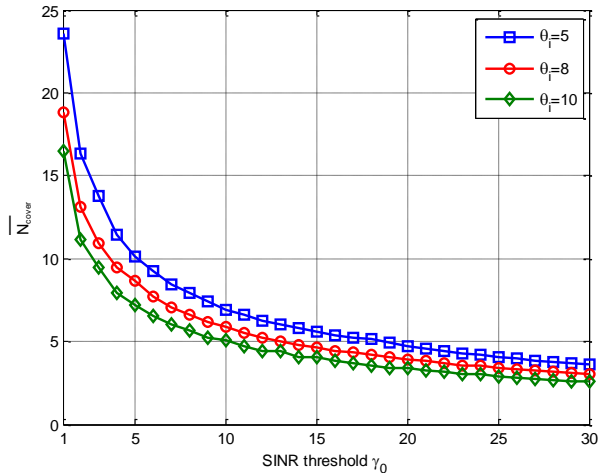}
\caption{\small  Average number of SBSs in a VC with respect to the SINR threshold ${\gamma _0}$  considering different ratios of the stationary user density to the SBS density ${\theta _i}$. }
\end{figure}

The effect of the SINR threshold  ${\gamma _0}$ and the ratio of the stationary user density to the SBS density  ${\theta _i}$ on the average number of SBSs in a VC $\overline {{N_{\mathrm{cover}}}} $ is evaluated in Fig. 6. Fig. 6 reveals that when  ${\gamma _0}$ is fixed $\overline {{N_{\mathrm{cover}}}} $  decreases with the increase of ${\theta _i}$. When ${\theta _i}$ is fixed, $\overline {{N_{\mathrm{cover}}}} $ decreases with the increase of the SINR threshold ${\gamma _0}$. Similar results are observed in \cite{Ge16}.

\begin{figure}[H]
\vspace{0.1in}
\centering
\includegraphics[width=8cm, draft=false]{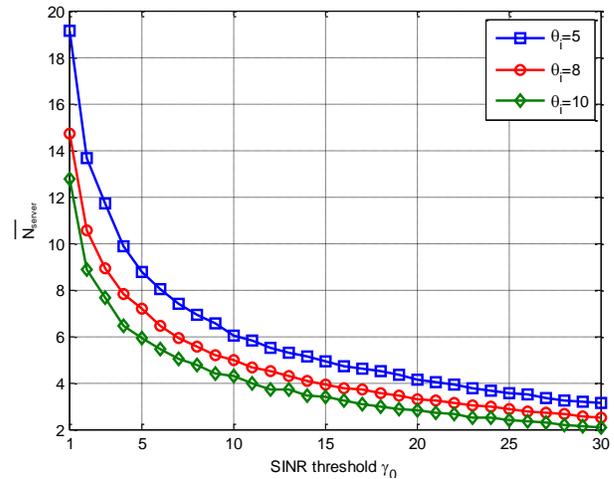}
\caption{\small  Average number of available SBSs in a VC with respect to the SINR threshold ${\gamma _0}$  considering different ratios of the stationary user density to the SBS density ${\theta _i}$. }
\end{figure}

The influence of the SINR threshold ${\gamma _0}$  and the ratio ${\theta _i}$  on the number of average available SBSs in a VC $\overline {{N_{\mathrm{server}}}} $  is shown in Fig. 7. As the figure shows, when ${\gamma _0}$  is fixed,  $\overline {{N_{\mathrm{server}}}} $  decreases with the increase of ${\theta _i}$. When ${\theta _i}$  is fixed, $\overline {{N_{\mathrm{server}}}} $  decreases with the increase of ${\gamma _0}$. By comparing Fig. 6 with Fig. 7,  we find that when ${\gamma _0}$  and ${\theta _i}$  are configured to be the same, $\overline {{N_{\mathrm{server}}}} $   in Fig. 6 is always smaller than $\overline {{N_{\mathrm{cover}}}} $  in Fig. 5, due to the limited service capability of SBSs.

\begin{figure}[H]
\vspace{0.1in}
\centering
\includegraphics[width=8cm, draft=false]{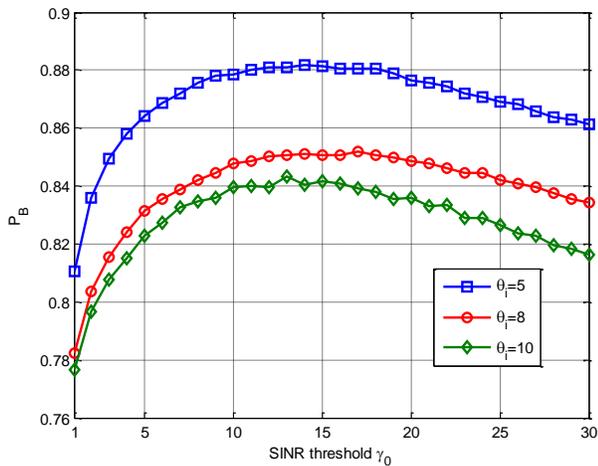}
\caption{\small  Blocking probability ${P_B}$  with respect to the SINR threshold  ${\gamma _0}$ considering different ratios of the stationary user density to the SBS density ${\theta _i}$. }
\end{figure}

The effect of the SINR threshold  ${\gamma _0}$ and the ratio ${\theta _i}$  on the blocking probability ${P_B}$  is shown in Fig. 8. When ${\gamma _0}$  is fixed, the blocking probability ${P_B}$   decreases with the increase of ${\theta _i}$. When ${\theta _i}$   is fixed, the blocking probability ${P_B}$   first increases with the increase of the SINR threshold ${\gamma _0}$. When the SINR threshold ${\gamma _0}$  is larger than a given threshold, the blocking probability ${P_B}$   decreases with the increase of the SINR threshold ${\gamma _0}$.

\begin{figure}[H]
\vspace{0.1in}
\centering
\includegraphics[width=8cm, draft=false]{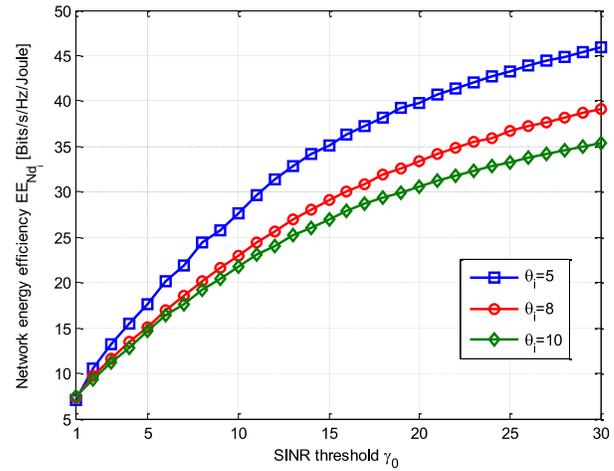}
\caption{\small  Network energy efficiency $E{E_{\mathrm{HS}_i}}$  with respect to the SINR threshold ${\gamma _0}$  considering different ratios of the stationary user density to the SBS density ${\theta _i}$. }
\end{figure}

The impact of the SINR threshold ${\gamma _0}$  and the ratio ${\theta _i}$  on the network energy efficiency is illustrated in Fig. 9. As the figure shows, when ${\theta _i}$ is fixed, the network energy efficiency  $E{E_{\mathrm{HS}_i}}$ increases with the increase of ${\gamma _0}$. Moreover, when the SINR threshold  ${\gamma _0}$ is fixed, the network energy efficiency $E{E_{\mathrm{HS}_i}}$  decreases with the increase of ${\theta _i}$.

\begin{figure}[H]
\vspace{0.1in}
\centering
\includegraphics[width=8cm, draft=false]{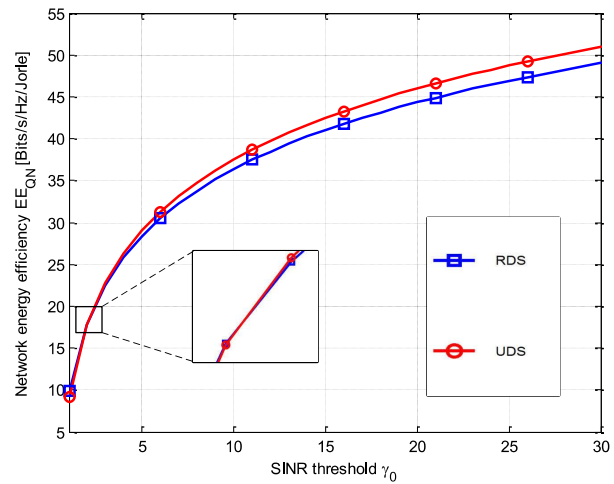}
\caption{\small  Network energy efficiency $E{E_{\mathrm{QN}}}$  with respect to the SINR threshold ${\gamma _0}$ for different SBS deploying strategies. }
\end{figure}

The impact of the SINR threshold ${\gamma _0}$  and the ratio ${\theta _i}$  on the energy efficiency of the small cell network contained in the $\mathrm{QN}$  system  is shown in Fig. 10. The user density based deploying strategy (UDS) is implemented by deploying 1000 SBSs into 10 hot spots based on the stationary number of users in each hot spot (e. g. the ratio of user density to base station density is the same for each hot spot), and the randomly deploying strategy (RDS) is implemented by letting SBSs and users be randomly and uniformly distributed in each hot spot without considering R\&A model. As the figure shows when $1 < {\gamma _0} < 2.2$, energy efficiency of the network based on the RDS is larger than that based on the UDS. When ${\gamma _0}>2.2$, the UDS based network energy efficiency becomes larger than the RDS based network energy efficiency.

\newpage
\section{Conclusion}

Based on the R\&A model, the impact of the non-uniformly distributed users on performance of 5G ultra-dense small cell networks is firstly investigated in this paper. Moreover, the queueing network theory is firstly adopted to extend the analysis of R\&A model to the temporal domain. Based on the theoretical analysis results for the stationary number of users, the coverage probability and the blocking probability are evaluated. Furthermore, by adopting virtual cell technology, the theoretical analyses of the throughput and the energy efficiency for each hot spot are proposed. By the simulations of the proposed model, we observe that when the virtual cell technology is adopted, the network energy efficiency is complicatedly coupled other than varying monotonically with the ratio of the stationary user density to the SBS density. This result provides insights into the important issues such as the optimal deployment of SBSs.

\section{Appendix}
\subsection{Coverage probability}

The Laplace transform of the random variable $Z$ is denoted by ${{\cal L}_Z}\left( \varepsilon  \right)$, where $Z$ is

\[Z = \frac{1}{{{\rho _0}{G_s}}} + \frac{{\sum\nolimits_{i = 0}^{{N_S}\left( i \right)\left( {1 - {P_{\mathrm{cover}}}} \right)} {{G_i}R_i^{ - {\alpha _p}}} }}{{{G_s}R_s^{ - {\alpha _p}}}}.\tag{41}\]

Thus, the Laplace transform ${{\cal L}_Z}\left( \varepsilon  \right)$ is derived as (42), where $\exp \left( \cdot \right)$  is the exponential function and ${\mathbb{E}_{{G_s},{G_i},{R_s},{R_i}}}\left( \cdot \right)$  is the expectation operations on random variables ${G_s}$, ${G_i}$, ${R_s}$ and ${R_i}$.

\begin{figure*}[!t]
\[{{\cal L}_Z}\left( \varepsilon  \right) = {\mathbb{E}_{{G_s},{G_i},{R_s},{R_i}}}\left( {\exp \left( { - \varepsilon \left( {\frac{1}{{{\rho _0}{G_s}}} + \frac{{\sum\nolimits_{i = 0}^{{N_S}\left( i \right)\left( {1 - {P_{\mathrm{cover}}}} \right)} {{G_i}R_i^{ - {\alpha _p}}} }}{{{G_s}R_s^{ - {\alpha _p}}}}} \right)} \right)} \right),\tag{42}\]
\end{figure*}

And the PDFs of ${G_s}$, ${G_i}$, ${R_s}$ and ${R_i}$ are

\[\left\{ \begin{array}{l}
{f_{{G_s}}}\left( x \right) = \eta \exp \left( { - \eta x} \right)\\
{f_{{G_i}}}\left( x \right) = \eta \exp \left( { - \eta x} \right)\\
{f_{{R_s}}}\left( x \right) = \frac{{4x}}{{\pi l_i^2}}\left( {\arccos \frac{x}{{2{l_i}}} - \frac{x}{{2{l_i}}}\sqrt {1 - \frac{{{x^2}}}{{4l_i^2}}} } \right)\\
{f_{{R_i}}}\left( x \right) = \frac{{4x}}{{\pi l_i^2}}\left( {\arccos \frac{x}{{2{l_i}}} - \frac{x}{{2{l_i}}}\sqrt {1 - \frac{{{x^2}}}{{4l_i^2}}} } \right)
\end{array} \right..\tag{43}\]

Based on (6)(42)(43), the Laplace transform of $Z$ is further derived by (43).

\begin{figure*}[!t]
\[\begin{array}{l}
{{\cal L}_Z}\left( \varepsilon  \right) = {\mathbb{E}_{{G_s},{G_i},{R_s},{R_i}}}\left( {\exp \left( { - \varepsilon \left( {\frac{1}{{{\rho _0}{G_s}}} + \frac{{\sum\nolimits_{i = 0}^{{N_S}\left( i \right)\left( {1 - {P_{\mathrm{cover}}}} \right)} {{G_i}R_i^{ - {\alpha _p}}} }}{{{G_s}R_s^{ - {\alpha _p}}}}} \right)} \right)} \right)\\
\quad \quad \;\;\; = {\mathbb{E}_{{G_s},{G_i},{R_s},{R_i}}}\left( {\exp \left( {\frac{{ - \varepsilon }}{{{\rho _0}{G_s}}}} \right)\prod\nolimits_{i = 0}^{{N_S}\left( i \right)\left( {1 - {P_{\mathrm{cover}}}} \right)} {\exp \left( {\frac{{ - \varepsilon {G_i}R_i^{ - {\alpha _p}}}}{{{G_s}R_s^{ - {\alpha _p}}}}} \right)} } \right)\\
\quad \quad \;\;\; = {\mathbb{E}_{{G_s}}}\left( {\exp \left( {\frac{{ - \varepsilon }}{{{\rho _0}{G_s}}}} \right) \cdot {{\left( {{\mathbb{E}_{{G_i},{R_s},{R_i}}}\left( {\exp \left( {\frac{{ - \varepsilon {G_i}R_i^{ - {\alpha _p}}}}{{{G_s}R_s^{ - {\alpha _p}}}}} \right)} \right)} \right)}^{{N_S}\left( i \right)\left( {1 - {P_{\mathrm{cover}}}} \right)}}} \right)\\
\quad \quad \;\;\; = \int\limits_0^\infty  {\exp \left( {\frac{{ - \varepsilon }}{{{\rho _0}{G_s}}}} \right)} \left( {\eta \exp \left( { - \eta {G_s}} \right)} \right) \cdot {\left( {{\mathbb{E}_{{G_i},{R_s},{R_i}}}\left( {\exp \left( {\frac{{ - \varepsilon {G_i}R_i^{ - {\alpha _p}}}}{{{G_s}R_s^{ - {\alpha _p}}}}} \right)} \right)} \right)^{\overline {{N_{\mathrm{int}}}} }}d{G_s}\\
\quad \quad \;\;\; = \int\limits_0^\infty  {\eta \exp \left( {\frac{{ - \varepsilon }}{{{\rho _0}{G_s}}} - \eta {G_s}} \right)} {\left( {{\mathbb{E}_{{G_i},{R_s},{R_i}}}\left( {\exp \left( {\frac{{ - \varepsilon {G_i}R_i^{ - {\alpha _p}}}}{{{G_s}R_s^{ - {\alpha _p}}}}} \right)} \right)} \right)^{\overline {{N_{\mathrm{int}}}} }}d{G_s},
\end{array}\tag{43}\]
\end{figure*}
with (44).
\begin{figure*}[!t]
\[\begin{array}{l}
{\mathbb{E}_{{G_i},{R_s},{R_i}}}\left( {\exp \left( {\frac{{ - \varepsilon {G_i}R_i^{ - {\alpha _p}}}}{{{G_s}R_s^{ - {\alpha _p}}}}} \right)} \right) = \int\limits_0^{2{l_0}} {\int\limits_0^{2{l_0}} {\int\limits_0^\infty  {\exp \left( {\frac{{ - \varepsilon {G_i}R_i^{ - {\alpha _p}}}}{{{G_s}R_s^{ - {\alpha _p}}}}} \right)} } } \cdot \\
\quad \quad \quad \quad \quad \quad \quad \quad \quad \quad \quad \;\;\,\quad \left( {{f_{{G_i}}}\left( {{G_i}} \right){f_{{R_s}}}\left( {{R_s}} \right){f_{{R_i}}}\left( {{R_i}} \right)} \right)d{G_i}d{R_s}d{R_i},
\end{array}\tag{44}\]
\end{figure*}

By substituting (42) (44) into (43), the Laplace transform in (43) becomes (45).
\begin{figure*}[!t]
{\small{
\[\begin{array}{l}
{{\cal L}_Z}\left( \varepsilon  \right) = {\mathbb{E}_{{G_0}}}\left( {\exp \left( { - \frac{\varepsilon }{{{\rho _0}{G_s}}}} \right)} \right.\left( {\int\limits_0^\infty  {\int\limits_0^{2{l_i}} {\int\limits_0^{2{l_i}} {\exp \left( {\frac{{ - \varepsilon {G_i}R_i^{ - {\alpha _p}}}}{{{G_s}R_s^{ - {\alpha _p}}}}} \right)\left( {\frac{{4{R_s}}}{{\pi l_i^2}}\left( {\arccos \frac{{{R_s}}}{{2{l_i}}} - \frac{{{R_s}}}{{2{l_i}}}\sqrt {1 - \frac{{{R_s}^2}}{{4l_i^2}}} } \right)} \right)} } } } \right.\cdot\\
\quad \quad \quad \;\left. {{{\left. {\left( {\frac{{4{R_i}}}{{\pi l_i^2}}\left( {\arccos \frac{{{R_i}}}{{2{l_i}}} - \frac{{{R_i}}}{{2{l_i}}}\sqrt {1 - \frac{{{R_i}^2}}{{4l_i^2}}} } \right)} \right)\left( {\eta \exp \left( { - \eta {G_i}} \right)} \right)d{G_i}d{R_s}d{R_i}} \right)}^{\overline {{N_{\mathrm{int}}}} }}} \right)\\
\quad \quad \;\; = \int\limits_0^\infty  {\left( {\eta \exp \left( { - \frac{\varepsilon }{{{\rho _0}{G_s}}} - \eta {G_s}} \right)} \right.\left( {\int\limits_0^\infty  {\int\limits_0^{2{l_i}} {\int\limits_0^{2{l_i}} {\exp \left( {\frac{{ - \varepsilon {G_i}R_i^{ - {\alpha _p}}}}{{{G_s}R_s^{ - {\alpha _p}}}}} \right)\left( {\frac{{4{R_s}}}{{\pi l_i^2}}\left( {\arccos \frac{{{R_s}}}{{2{l_i}}} - \frac{{{R_s}}}{{2{l_i}}}\sqrt {1 - \frac{{{R_s}^2}}{{4l_i^2}}} } \right)} \right)} } } } \right.} \cdot\\
\quad \quad \quad \;\left. {{{\left. {\left( {\frac{{4{R_i}}}{{\pi l_i^2}}\left( {\arccos \frac{{{R_i}}}{{2{l_i}}} - \frac{{{R_i}}}{{2{l_i}}}\sqrt {1 - \frac{{{R_i}^2}}{{4l_i^2}}} } \right)} \right)\left( {\eta \exp \left( { - \eta {G_i}} \right)} \right)d{G_i}d{R_s}d{R_i}} \right)}^{\overline {{N_{\mathrm{int}}}} }}} \right)d{G_s}.
\end{array}\normalsize{\tag{45}}\]
}}
\end{figure*}

Similar to \cite{Abate95}, by the Euler summation on (45), ${P_{\mathrm{cover}}}$ becomes

\[\begin{array}{l}
{P_{{\rm{cover}}}} = {2^{ - {B_e}}}{\gamma _0}\exp \left( {\frac{{{A_e}}}{2}} \right)\sum\limits_{{b_e} = 0}^{{B_e}} {\left( \begin{array}{l}
{B_e}\\
{b_e}
\end{array} \right)}  \times \\
\quad \quad \quad \; \sum\limits_{{c_e} = 0}^{{C_e} + {b_e}} {\frac{{{{\left( { - 1} \right)}^{{c_e}}}}}{{{D_e}}}{\mathop{\rm Re}\nolimits} \left( {\frac{{{{\cal L}_Z}\left( \hbar  \right)}}{\hbar }} \right)} \quad .
\end{array}\tag{46}\]

Based on the analysis in \cite{OCinneide97}, ${A_e}$, ${B_e}$ and ${C_e}$ should be no less than $t\ln 10$, $1.243t - 1$  and $1.467t$  respectively to obtain a numerical accuracy of ${10^{ - t}}$. In this paper, ${A_e}$, ${B_e}$  and ${C_e}$ are configured to be $8\ln 10$, $11$  and $14$  in order to obtain a numerical accuracy of ${10^{ - 9}}$  for the theoretical analysis. The parameter $\hbar $ is expressed as $\hbar  = \left( {{A_e} + 2\pi {c_e}J} \right)/\left( {2\gamma _0^{ - 1}} \right)$, where $J$  is the imaginary unit and ${\mathop{\rm Re}\limits} \left( \cdot \right)$  denotes the real part of the given variable in parentheses. The parameter ${D_e}$  is

\[{D_e} = \left\{ \begin{array}{l}
2\quad \mathrm{if}\,{c_e} = 0\\
1\quad \,\mathrm{others}
\end{array} \right..\tag{47}\]

By substituting (45) into (46), the coverage probability ${P_{\mathrm{cover}}}$ becomes (12).


\newpage
\end{document}